\documentclass{sig-alternate-05-2015}

\usepackage{url}
\usepackage{amsmath}
\usepackage{graphicx,subfigure}
\usepackage[colorlinks,linkcolor=red,citecolor=blue,urlcolor=blue]{hyperref}

\usepackage[sort&compress,numbers]{natbib}
\usepackage[flushleft]{threeparttable}

\usepackage{algorithm}\usepackage{algpseudocode}

\usepackage{tabularx}
\usepackage{mdwlist}

\usepackage{multirow}
\usepackage{rotating}
\input{Definitions}

\newcommand\rbracket[1]{\left(#1\right)}
\newcommand\cbracket[1]{\left\{#1\right\}}

\DeclareMathOperator{\abf}{\mathbf{a}}

\DeclareMathOperator{\alphabf}{\boldsymbol{\alpha}}
\DeclareMathOperator{\cbf}{\mathbf{c}}
\DeclareMathOperator{\Mbf}{\mathbf{M}}
\DeclareMathOperator{\wbf}{\mathbf{w}}

\DeclareMathOperator{\wbfhat}{\mathbf{\hat{w}}}
\DeclareMathOperator{\wbftilde}{\mathbf{\tilde{w}}}

\DeclareMathOperator{\Naturals}{\mathbb{N}}

\algtext*{EndFunction}
\algtext*{EndWhile}
\algtext*{EndFor}
\algtext*{EndIf}

\begin{document}

\conferenceinfo{RecSys 2016}{September 15--19, 2016, Boston, MA}

\title{Adaptive, Personalized Diversity for Visual Discovery}

\numberofauthors{7}
\author{
  \alignauthor Choon Hui Teo\\
  \affaddr{Amazon}\\            
  \email{choonhui@amazon.com}
  \alignauthor Houssam Nassif\\
  \affaddr{Amazon}\\
  \email{houssamn@amazon.com}
  \alignauthor Daniel Hill\\
  \affaddr{Amazon}\\
  \email{daniehil@amazon.com}
  \and
  \alignauthor Sriram Srinavasan\\
  \affaddr{UC Santa Cruz}
  \email{ssriniv9@ucsc.edu}
  \alignauthor Mitchell Goodman\\
  \affaddr{Amazon}\\
  \email{migood@amazon.com}
  \alignauthor Vijai Mohan\\
  \affaddr{Amazon}\\
  \email{vijaim@amazon.com}
  \and 
  \alignauthor S. V. N. Vishwanathan\\
  \affaddr{Amazon \& UC Santa Cruz}\\
  \email{vishy@amazon.com}
}

\CopyrightYear{2016} 
\setcopyright{rightsretained} 
\conferenceinfo{RecSys '16}{September 15-19, 2016, Boston , MA, USA} 
\isbn{978-1-4503-4035-9/16/09}
\doi{http://dx.doi.org/10.1145/2959100.2959171}

\maketitle
\begin{abstract}
  Search queries are appropriate when users have explicit intent, but
  they perform poorly when the intent is difficult to express or if
  the user is simply looking to be inspired. Visual browsing systems
  allow e-commerce platforms to address these scenarios while offering
  the user an engaging shopping experience. Here we explore extensions
  in the direction of adaptive personalization and item
  diversification within Stream, a new form of visual browsing and
  discovery by Amazon. Our system presents the user with a diverse set
  of interesting items while adapting to user interactions. Our
  solution consists of three components (1) a Bayesian regression
  model for scoring the relevance of items while leveraging
  uncertainty, (2) a submodular diversification framework that
  re-ranks the top scoring items based on category, and (3)
  personalized category preferences learned from the user's
  behavior. When tested on live traffic, our algorithms show a strong
  lift in click-through-rate and session duration.
\end{abstract}

%
%
\printccsdesc
\keywords{Machine Learning; Submodular Functions; Diversity;
  Personalization; Explore-Exploit; Multi-Armed Bandits}

\section{Introduction}
\label{sec:Introduction}

The brick-and-mortar shopping experience is characterized by visual
browsing where the user is able to quickly scan a large number of
potential purchases. The user has a high potential to discover new
items while maintaining the ability to focus attention on items of
particular interest. This problem of discoverability is more
challenging in e-commerce where it can be difficult to expose the
entirety of an online retailer's catalog. The in-store browsing
experience is not well-replicated by search engines that restrict item
discovery to items relevant to an explicit search query.  Therefore,
an online visual browsing experience may greatly aid users in item
discovery.

\begin{figure}[h]
  \begin{center}
    \includegraphics[width=1.0\linewidth]{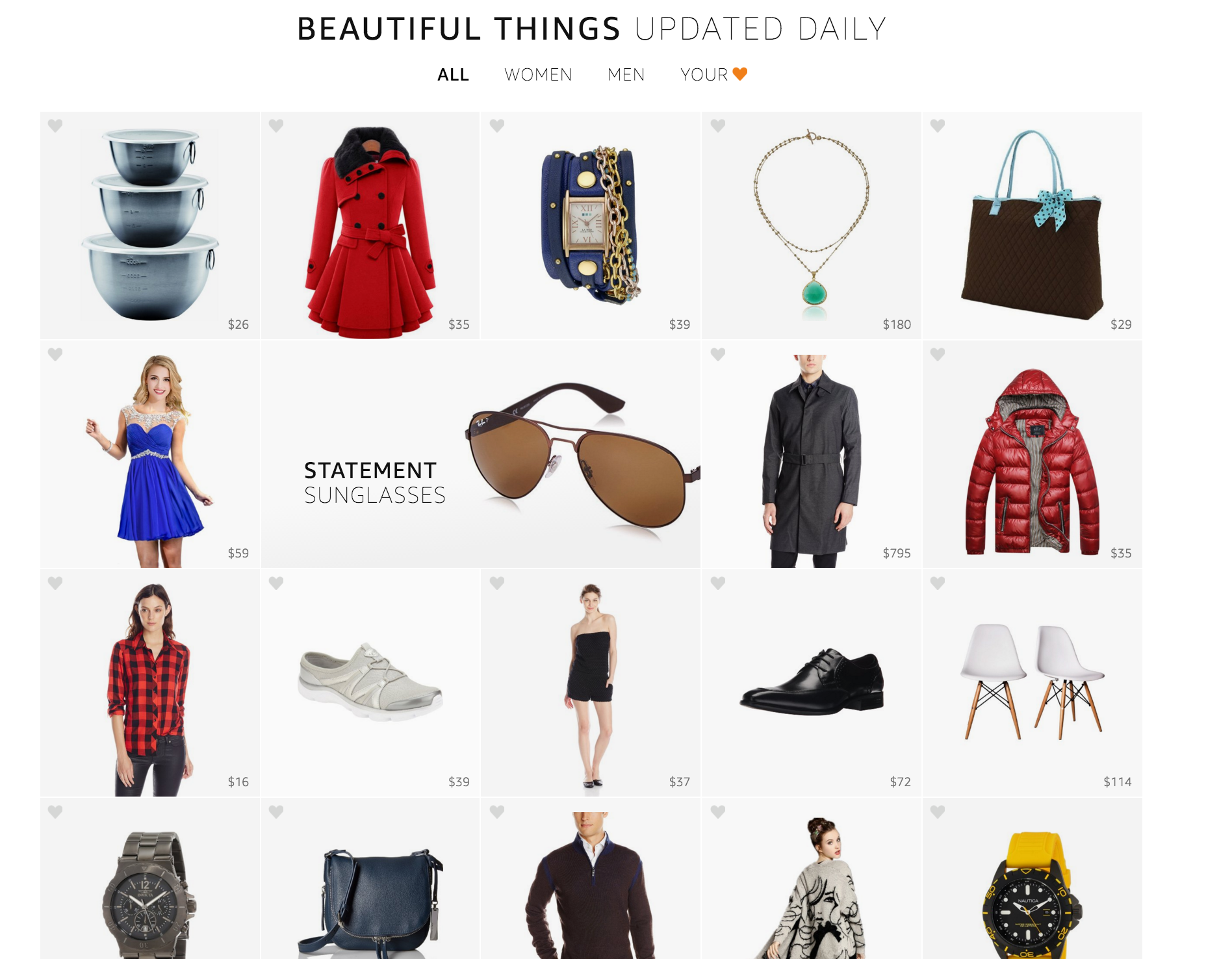}
    \end{center}
  \caption{Screenshot of \href{http://www.amazon.com/stream}{www.amazon.com/stream}.}
  \label{fig:screenshot}
\end{figure}

One effort in this direction is Amazon Stream
(figure~\ref{fig:screenshot}), a new website for fashion discovery
developed by Amazon
(\href{http://www.amazon.com/stream}{www.amazon.com/stream}). Stream
enables users to easily discover popular, new, and relevant
fashionable items without the need for search queries or to sieve
through less relevant items. Toward this end, we have built a
personalizable system that ranks and diversifies items scored by an
explore-exploit algorithm. These items are presented to the user as an
infinite scroll where each item can be interacted with by
clicking. This paper outlines elements of our diverse and personalized
visual shopping experience approach.

\section{Scoring item relevance}
\label{sec:CandidateScoring}

As the first step in generating the stream, we score each item in the
Stream catalog for relevance to our users. We use click probability $P(click ~|~ item~ is~ viewed)$ to quantify relevance. \textit{Click} refers to any of the following activities: save item as
favorite, visit the item's detail page, or open the modal window for
the item.

We learn this probability distribution by using a Bayesian Linear
Probit regression~\cite{graepel2010web} that maps item attributes onto
click probabilities. Item attributes are represented by categorical
variables. They include product features such as brand, color,
department, price, size, as well as a unique identifier for each item.
The regression model learns a set of weights where each weight
represents the contribution of a single attribute value to the click
probability.  When evaluating an item, we say that a weight $x_i$ is
active if its corresponding attribute is present in the item.  Thus,
the expected click probability of an item is given by
\begin{equation}
  \label{eq:blip}
P(click ~|~ item~ is~ viewed ) = \Phi^{-1}(\sum_{x_i~ is~ active}x_i)
\end{equation}
where $\Phi^{-1}(\cdot)$ is the probit function. We assume model
weights are generated independently from a prior Normal distribution.
As we observe user actions, we update the active weights posterior distributions accordingly~\cite{graepel2010web}.

Our catalog consists of hundreds of thousands of items with thousands
of items added or purged daily. The expected click probability may
unduly favor popular items over underexposed or recently added
items. This is because underexposed items will likely be associated
with undertrained model weights that potentially underestimate the
true click probability. This poses an exploration-exploitation
dilemma: balancing exploiting highly relevant items with exploring undersampled sections of the catalog. Multi-armed bandit algorithms such as Thompson sampling \cite{thompson1933likelihood} and Upper Confidence Bound \cite{auer2003using} solve this problem. Here we use a Thompson
sampling approach to score items as it performs well under delayed
feedback~\cite{chapelle2011empirical}.

We apply Thompson sampling by sampling each model weight
from its posterior distribution conditioned on the training data. We
then use the sampled weights to compute equation~\eq{eq:blip} for each item. Under this procedure, an item will receive a score close to
its expected click probability unless its attributes are associated
with highly uncertain model weights. Therefore, the items with the
highest scores will be a mix of popular products with high expected
values (an exploitation strategy) and underexposed products that were
randomly scored high above their expected value (an exploration
strategy).

Finally, we note that the customer controls the number of products
viewed via scrolling. For simplicity we have omitted a model of
customer browsing that predicts the number of views per session.
However, such a model would allow us to estimate the expected clicks
per session directly rather than the expected clicks per item viewed.

\section{Submodular Diversification}
\label{sec:Diversification}
We now have a set of items that are scored based on both their
relevancy and exposure.  Similar items tend to be assigned similar
scores, so ranking these items by their scores creates a list where
highly similar items are clustered together. However, our user
experience studies indicate that a diverse item list helps users
discover relevant items faster, as it exposes more variety in a fixed
number of slots~\cite{nassif2016music}. Similar to cases in web search
\cite{agrawal2009diversifying}, blog posts \cite{el2009turning}, news articles \cite{ahmed2012fair} and music recommendation~\cite{nassif2016music}, we apply diversification to
alleviate information redundancy.

We formulate the selection and ranking of a diverse subset of items as
a submodular optimization problem~\cite{fujishige2005submodular}. Let
$\Acal := \cbr{\ab_{1}, \ldots, \ab_{n}}$ denote the attribute vectors
for a set of $n$ items. All attribute vectors are one-hot encoded in
$d$-dimensions such that $\ab_i \in \cbracket{0,1}^d$. Let $\wb$
denote a $d$-dimensional vector which encodes user preferences for
each of the $d$ item attributes. In addition, let $s(\ab)$ denote a
function which maps an item to a real-valued utility score such as the
click probability.  A submodular objective function $\rho$ for
selecting $k$ items from $\Acal$ is given by
\begin{equation}
  \label{eq:submod_fn}
  \rho(\Acal_{k}, \wb) = \inner{\wb}{\log \rbr{1 + \sum_{\ab_i \in \Acal_{k}}
  \ab_{i}}} + \sum_{a_i \in \Acal_{k}} s(\ab_i),
\end{equation}
where $\Acal_{k}$ is a subset of $\Acal$ of size $k$. We use the
convention of applying the logarithm to a vector's individual
components.  The optimal subset is given by
\begin{equation}
  \label{eq:submod_obj}                                                                                               
 \Acal^{*}_k := \argmax\limits_{\Acal_{k} \subseteq \Acal, |\Acal_{k}|
  = k} \rho(\Acal_{k}, \wb).
\end{equation}

In equation \eq{eq:submod_fn}, the weight $w_j$ emphasizes the
importance of the $j^{th}$ attribute to a user. By summing over
$\ab_{i}$ we obtain the number of times that attribute occurs in the
collection $\Acal_{k}$.  However, the utility of each attribute does
not grow linearly with its count. Instead, the logarithmic term
ensures that the incremental utility of the $j^{th}$ attribute
diminishes as the number of items with $j$ attributes increases in
$\Acal_{k}$. In other words, if a user loves shoes, showing only shoes
does not lead to a good user experience. The scale of
$s(\ab)$ can be tuned to adjust the relative importance of the item's
utility.

Equation \eq{eq:submod_obj} is a special case of
the NP-hard maximum set cover problem. Nevertheless, we can use an iterative
greedy procedure to obtain a near-optimal solution~\cite{nemhauser1978analysis}:
\begin{equation}
  \label{eq:greedy}
  \Acal_0 := \emptyset ~\text{and}~ 
  \Acal_{i+1} := \Acal_i \cup \cbr{\argmax\limits_{\ab \in \Acal \backslash
      A_i} \rho(\Acal_i \cup \cbr{\ab}, \wb)},
\end{equation}
with a runtime complexity of $O(dkn)$. We further boost efficiency by using
the CELF lazy evaluation algorithm~\cite{leskovec2007cost, el2009turning}.

We also optimize the diversification process by transforming the
attribute vector space into a single categorical space. This is
achieved by mapping each possible attribute vector to a unique
category. We restrict the size of this categorical space by coarsening
our representation of attributes, such as price, and eliminating
attributes that do not need to be diversified, such as average rating
in customer reviews. The final set for diversification includes on the
order of 100 mutually exclusive categories.

\section{Personalization}
\label{sec:Personalization}
Category weights $\wbf$ in equation~\eq{eq:submod_fn} control the trade-off
between item utility and category popularity for selecting and
ranking items.  Unlike methods such as
\cite{yue2011submodularBandits} and \cite{ahmed2012fair}, we do not
optimize these weights directly. This allows us to exploit user
behavioral data that was not subject to diversification. In the
following sections, we describe the procedure for learning global and
personalized category weights.

\subsection{Learning Adaptive Global Weights}
\label{sec:LearnGlobWeights}
Though items in the stream are displayed in a grid layout
(figure~\ref{fig:screenshot}), we assume that users scan them
linearly from left to right and top to bottom. Hence, we can leverage
existing work on click modeling with linear positional bias
correction.

To learn the weights $\wbf$ in equation \eq{eq:submod_fn}, we compared
logistic regression, clicks over expected clicks~\cite{zhang2007comparing}, and click-thru-rate
(CTR) with additive smoothing. Each model was tested on several weeks
of historical click logs. Even though the CTR approach does not
include corrections for position bias, we found that the three methods
gave similar results. We therefore opted to use CTR with additive
smoothing as it is simple to implement yet provides additional
controls to handle update cycle fluctuations.

Let $c_{j}$ and $v_{j}$ represent the number of clicks and views,
respectively, for items from the $j^{th}$ category.  Under the CTR
with additive smoothing model, the weight for the $j^{th}$ category
$w_j$ is estimated as:
\begin{equation}
  \label{eq:smoothed_Ctr}
  w_j = \frac{c_{j} + \alpha}{v_{j} + \alpha + \beta},
\end{equation}
where $\alpha$ and $\beta$ are priors that enforce minimum category
weights. $\alpha$ and $\beta$ can be tuned
based on traffic volume to avoid dramatic fluctuations in update
intervals. In fact, one can interpret $w_j$ as distributed according
to a Binomial distribution whose parameter $p_j$ is drawn from a Beta
distribution with parameters $\alpha$ and $\beta$ estimated from data.

\subsection{Learning User Specific Weights}
\label{sec:LearningUserSpecific}

The adaptively tuned category weights described in the previous
section were derived from the aggregate clicks of all users. They capture the general popularity of the categories but not the
interests of particular users. To personalize the presenation of
items and categories, we need to first build a user model. Unlike some
personalized applications where user feedback is immediate and
irreversible, our model must handle delayed feedback and react quickly
and efficiently to \textit{evolving} preferences where a positive
indicator of interest may no longer be predictive at a later date.

\subsubsection{User Modeling}
\label{sec:user_model}

User modeling methods such as matrix factorization
\cite{koren2009matrix} and hierarchical Bayesian models
\cite{low2011multiple} are not directly applicable in our case because
these methods work best with a massive collection of data from all
users, and are thus less sensitive to changing patterns in a small
fraction of training records. Instead, we model the $u^{th}$ user's
level of interest in the $d$ categories as a random vector $\wbf_u$
drawn from a Dirichlet distribution parameterized by $\alphabf_0$.
The distribution of a user's clicks on each category is denoted as
$\cbf_u$ and is modeled as a random variable drawn from a Multinomial
distribution parameterized by the interest vector and the user's total
number of clicks $m_u$:
\begin{align}
  \wbf_u & \sim \text{Dirichlet}(\alphabf_0), \label{eq:user_interest} \\
  \cbf_u & \sim \text{Multinomial}(\wbf_u, m_u). \label{eq:user_clicks}
\end{align}

The posterior distribution of the user interest vector $\wbf_u$ is given proportionally by 
\begin{equation}
  \label{eq:user_interest_posterior}
  P(\wbf_u|\cbf_u,\alphabf_0) \propto P(\cbf_u|\wbf_u)P(\wbf_u|\alphabf_0).
\end{equation}
As Dirichlet is conjugate prior for the Multinomial distribution,  
this posterior is also Dirichlet with mean $\wbfhat_u$:
\begin{equation}
  \label{eq:est_user_vector}
  \wbfhat_u = (\cbf_u+\alphabf_0)\norm{\cbf_u+\alphabf_0}_{1}^{-1}.
\end{equation}
We estimate the personalized weight vector $\wbf_u$ from the user's
past clicks $\cbf_u = (c_{u1},\ldots,c_{ud})$ where
$c_{uj} \in \Naturals$ is the number of clicks for category $j$. Dirichlet prior parameter vector $\alphabf_0$ can be chosen
to match pre-specified business rules or to highlight certain category
preferences.

In addition to being simple to implement, this Multinomial-Dirichlet
user model allows straightforward incorporation of different types of
user signals by adding them as counts to $\cbf_u$. Similarly, evolving
preferences are handled by deducting older signals from
$\cbf_u$. Finally, the personalized diversification can be achieved by
plugging the estimated user interest vector $\wbfhat_u$ into
equation~\eq{eq:submod_fn}.

\subsubsection{User Click Signal Diffusion}
\label{sec:over_p13n}
User interests are correlated between different categories. For
example, a male interested in high end fashion is likely interested in
both suits and expensive watches. This concept is especially useful
when users have concentrated clicks on only a few categories, which
would otherwise tend to inhibit diversified recommendations.

To ensure a user is exposed to related categories, we diffuse our
Dirichlet updates between categories. In addition to a user's click
incrementing the item's specific category count, the update is
diffused across \textit{related} categories. Let $\Mbf$ be a
$d$-by-$d$ matrix where the entry $M_{ij}$ is the ratio of the number
of users who clicked items from both categories $i$ and $j$ to the
number of users who clicked items from category $j$.  To alleviate
noise, we use only the top categories each user has actively
interacted with. The \textit{smoothed} interest vector $\wbftilde_u$
of the user interest vector $\wbfhat_u$ of equation
\eq{eq:est_user_vector} becomes:
\begin{equation}
  \label{eq:smoothed_user_vector}
  \wbftilde_u = \Mbf\wbfhat_u \norm{\Mbf\wbfhat_u}_{1}^{-1}.
\end{equation}

There are several common approaches for quantifying the pairwise
association between two categories, such as pointwise mutual
information or Pearson's correlation coefficients. We use the
asymmetric conditional probability that ``users who like category $i$
also like category $j$'' because we have observed such user behavior
patterns in our data. Moreover, the leading eigenvector of the
co-interest probability matrix $\Mbf$ can be seen as a global interest
vector, towards which equation \eq{eq:smoothed_user_vector} smooths
the user interest vector $\wbfhat_u$.

\section{Experiments}
\label{sec:Experiments}

We conducted several online experiments to test the effectiveness of
the methods presented above. We evaluated the treatment impact on
engagement by calculating session duration, number of items viewed,
and total click-through-rate. Each experiment lasted at least one week
and involved at least 100,000 users per group. Some statistics are
reported for population subsets of at least 10,000 users per
group. Statistical significance was determined via Welch's
t-test. Table~\ref{table:results} shows experimental results. Columns
should not be compared directly to each other, as each experiment's
control is different. Columns confer incremental improvements on top
of the previous treatment.

\begin{table*}[t]
  \begin{center}
    \begin{tabular}{|l|c|c|c|} \hline
      & Submodular diversifier
      & Adaptive weights 
      & Personalized weights \\ \hline
      Duration  &           0.05\%    &  ~  \textbf{5.39\%}    & ~~~           1.10\%     \\ \hline
      Views     &  \textbf{-1.32\%}   &  ~  \textbf{1.08\%}    & ~~~  \textbf{-4.95\%}    \\ \hline
      CTR       &  \textbf{ 9.82\%}   &  ~  \textbf{8.29\%}    & ~~~  \textbf{12.58\%}    \\ \hline
    \end{tabular}
    \caption{Experimental results evaluated using per-session
      engagement metrics: duration, aggregated view count, and
      aggregated click-through-rate. Column 2 shows incremental
      changes between a \textit{submodular vs multinomial}
      diversifier. Column 3 shows incremental changes of
      \textit{adaptive global category weights vs static manual weights}. 
       Column 4 shows incremental changes for
      \textit{personalized category weights vs global weights}. Highlighted
      results are statistically significant at the 0.05 level.}
    \label{table:results}
  \end{center}
\end{table*}

\subsection{Submodular Diversification}
\label{sec:expt_div}
We first compared our submodular diversifier of
Section~\ref{sec:Diversification} to a simpler approach where
categorical weights and item scores are considered separately. The
model of section~\ref{sec:CandidateScoring} assigns a score to each
item. Within each category, we rank the items in descending order. We
calculate a multinomial distribution over item categories using global
click propensities. For each slot in the stream, we first sample a
category from the multinomial, and then fill it with that category's
highest scoring item.

Experimental results (Table~\ref{table:results}, column 2) show that,
within the same session duration, the amount of items viewed by the
users subjected to submodular diversification decreased by $1.32\%$
while CTR increased by $9.82\%$. The submodular diversifier produces a
more diverse stream as categories are penalized more aggressively
for being shown repeatedly. These results are consistent with our
expectation of users being more interested in diversified streams.

\subsection{Adaptive Category Weights}
\label{sec:expt_adaptive_weights}
In a second experiment, we tested the impact of learning adaptive
category weights from global click propensities as described in
Section~\ref{sec:LearnGlobWeights}. We compared a submodular
diversifier that uses the adaptive weights, versus one that uses a set
of weights chosen manually by Amazon fashion experts. Experimental
results (Table~\ref{table:results}, column 3) show that adaptive
weights greatly out-performed manual weights. CTR increased by
$8.29\%$ and session duration increased by $5.39\%$. This indicates
that users are highly sensitive to the categorical mix of their
experience.

\subsection{Personalization}
\label{sec:expt_p13n}
In our final experiment, we exposed users to category weights that
were adapted to their own click behavior, as described in
Section~\ref{sec:LearningUserSpecific}. These weights were calculated
for users whose history included at least 10 clicks, so this
experiment only applied to the most active users.  In the control
group, users were exposed to streams built using non-personalized
weights. Experimental results (Table~\ref{table:results}, column 4)
show that personalized weights yielded a $12.58\%$ increase in CTR
with almost no impact on duration. Interestingly, the amount of items
that were viewed in the stream decreased by $4.95\%$. This indicates
that users with personalized diversity weights had a more efficient
experience where they interacted with more items while scrolling
through less.

\section{Conclusion}
\label{sec:Conclusion}
We present Amazon Stream, a visual browsing system that
emphasizes adaptive and personalized diversity in user
experience. Such methods can be applied to any online system where
the user seeks to discover content in the absence of an explicit
search query.  Our experimental results show the tangible incremental
impact on key engagement metrics of submodular diversification,
adaptive global weights, and personalization.

\scriptsize{
\paragraph{Acknowledgments}
The authors thank Charles Elkan, Matthias Seeger, and the anonymous
reviewers for their helpful comments.
}

\bibliographystyle{abbrv}
\scriptsize{
\begingroup
\raggedright
\bibliography{recsys2016}
\endgroup
}
\end{document}